\documentclass{svproc}
\usepackage{url}

\usepackage{amsmath}
\usepackage{graphicx}
\usepackage{subcaption}
\graphicspath{}
\usepackage{multicol}
\usepackage{algorithmicx, algpseudocode}
\usepackage{float}
\usepackage{stfloats}
\usepackage{url}
\usepackage{wrapfig}
\usepackage{color}
\usepackage[dvipsnames]{xcolor}
\usepackage{hyperref}
\usepackage{cite}
\usepackage{multirow}
\usepackage{enumitem}

\begin{document}
\mainmatter              
\title{Novel Edge and Density Metrics for Link Cohesion}
\titlerunning{Link Cohesion}  
%
\author{Cetin Savkli \and Catherine Schwartz \and
Amanda Galante \and Jonathan Cohen}
\authorrunning{Cetin Savkli et al.} 
%
\tocauthor{Cetin Savkli, Catherine Schwartz, Amanda Galante, and Jonathan Cohen}
\institute{Johns Hopkins University Applied Physics Laboratory, Laurel MD 20723, USA \\
\email{amanda.galante@jhuapl.edu}}

\maketitle              

\begin{abstract}
We present a new metric of link cohesion for measuring the strength of edges in complex, highly connected graphs.   
Link cohesion accounts for local small hop connections and associated node degrees and can be used to support edge scoring and graph simplification.  
We also present a novel graph density measure to estimate the average cohesion across nodes.  
Link cohesion and the density measure are employed to demonstrate community detection through graph sparsification by maximizing graph density.  
Link cohesion is also shown to be loosely correlated with edge betweenness centrality.  
\keywords{link cohesion, graph sparsification, graph density, centrality, community detection}
\end{abstract}
%




\section{Introduction}\label{sec:intro}

Real world networks, or graphs defined as a set of nodes and connecting edges, tend to be highly connected, particularly in the cyber domain.  Analysis of these large, highly connected networks is generally computationally expensive.  While many techniques to analyze graphs have been developed (e.g., \cite{freeman1977set, marsden2002egocentric}), there are very few metrics that score edges themselves, e.g., edge betweenness \cite{girvan2002community}.  We posit that edge scores, calculated from local graph properties, can be used to help reduce complexity and analysis of large graphs. 

One common graph analysis technique is community detection, where connected nodes are clustered into communities.  Community detection techniques have been applied to social, mobile phone, biological, and legislative networks, to name a few \cite{porter2009communities}.  Community detection is known to be challenging unless edges are sparse \cite{fortunato2010community}.  This is particularly true for real-world graphs, which tend to become more dense with increasing degree and decreasing diameters over time \cite{leskovec2007graph}.  In light of this, as real-world graphs become larger and more interconnected, community detection and other graph analysis become more challenging and it is critical to identify techniques to help manage analysis of large graphs.  One approach, graph sparsification through edge removals, has been used to simplify networks for community detection \cite{satuluri2011local}; we seek to further explore the advantages of sparsification using a new concept we refer to as link cohesion. 


The main contributions of this paper are i) a new local edge cohesion calculation to support edge scoring and graph simplification and ii) an automated pruning approach to sparsify the graph by optimizing the newly developed density metric that leverages link cohesion. The link cohesion of an edge is a new concept that considers 1-, 2-, and 3-hop paths connecting end points while also taking into account the degree of nodes involved on those paths. In the following sections, we provide a description of the new metrics, their potential uses, and accompanying results that demonstrate the value of link-cohesion-based pruning in detecting communities on real world and synthetic data sets.




\section{Link Cohesion Calculation}\label{sec:method} 

The expected number of edges between two nodes $i$ and $j$ is given by $k_i k_j / 2|E|$, where $|E|$ is the number of edges in the graph and $k_i$ is the degree of node $i$.  In a undirected graph with only one edge allowed between any two nodes, this is equivalent to the likelihood of the edge.  This likelihood concept can be expanded to alternate paths created by triangles and quadrilaterals about each edge.  
We posit that edges are more cohesive when they are supported by alternate paths and that the value of those alternate paths is inversely proportional to their likelihood.  Using these concepts, we define \emph{link cohesion} as a means to assess how supported an edge is relative to other edges, accounting for the number of nearby alternate paths and associated node degrees.  We do this by calculating a score using 1-, 2-, and 3-hop connections, where supporting links are valued based on their inverse likelihood.  That is, links with higher likelihood and therefore a lower link cohesion score are presumed to provide less value to local connections.


The calculation for link cohesion is as follows:

%
\noindent 1. \textbf{Calculate hop-based link strengths}

Let $k_i$ denote the degree of each vertex $i$.  Consider 1-hop, 2-hop, and 3-hop connections between connected nodes $i$ and $j$.  

\hspace{0.2cm} The single link strength $a_{1,ij}$ of edge $e_{ij}$ for each direct connection will be measured as:
\begin{equation*}
a_{1,ij} = \frac{1}{k_i k_j}
\end{equation*}
Note that this is proportional to the inverse of the expected number edges between nodes $i$ and $j$, $k_i k_j / 2|E|$.  In this way, the single link strength places greater strength on links with a lower likelihood.

\hspace{0.2cm} The double link strength $a_{2,ij}$ of edge $e_{ij}$ for each direct connection with additional 2-hop connections will be measured as:
\begin{equation*}
a_{2, ij} = \frac{1}{(k_i k_j)^2}\sum_{\text{2-hop paths thru node $l$ from } e_{ij}} {\frac{1}{k_l^2}}
\end{equation*}
This calculation is similar to single link evaluation in that it penalizes triangles by the degree of the corners that form the triangle and places greater strength on links supported by triangles with lower likelihood.  The squared term is due to the fact that each node participates in two links.

\hspace{0.2cm} Similarly, the triple link strength $a_{3,ij}$ of edge $e_{ij}$ for each direct connection with additional 3-hop connections will be measured as:
\begin{equation*}
a_{3, ij} = \frac{1}{(k_i k_j)^2}\sum_{\text{Unique 3-hop paths thru nodes $m, n$ from } e_{ij}} {\frac{1}{(k_m k_n)^2}}
\end{equation*}

\begin{wrapfigure}{r}{0.3\textwidth}
	\centering
	\includegraphics[width = 1.5in]{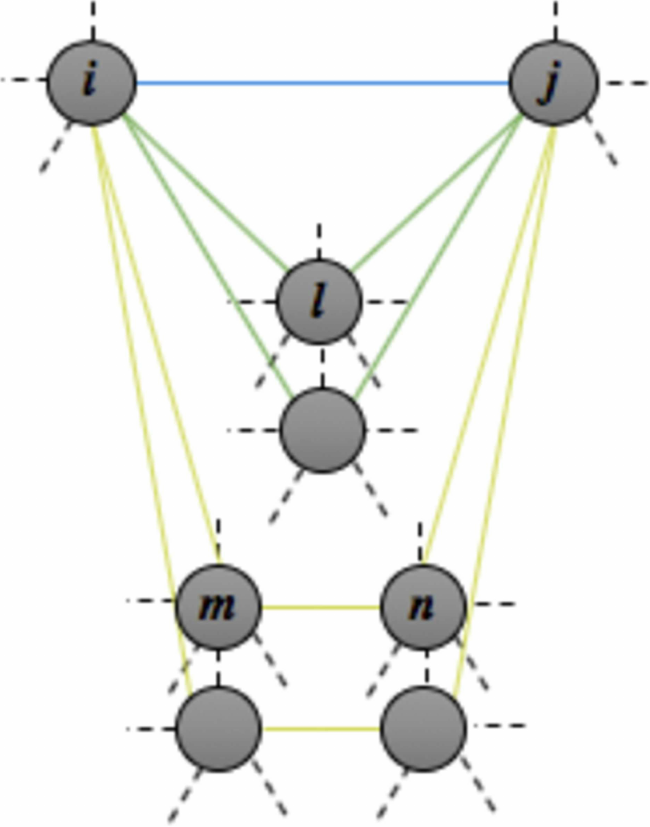}
	\caption{\label{fig:hops} Link cohesion uses 1-hop, 2-hop, and 3-hop connections}  
\end{wrapfigure}

In carrying out this summation, we ensure that there are no circular loops between pairs of nodes, following only 3-hop paths as shown by the path through nodes $m$ and $n$ in Figure~\ref{fig:hops}.  While this third measurement involves 3-hop paths, calculation needs to only be performed for pairs in which a link exists by re-using the 2-hop result.  
This leads to a significant decrease in the number of computations that are needed for 3-hop links.
The complexity this algorithm, as described, is $\sum_i (k_i)^2$; however, in implementing it by binning edges based on low-degree adjacent vertices allows the 3-cycle and 4-cycle contributions to be enumerated with worst case work of $|E|^{3/2}$ \cite{cohen2019trapeze}.

\noindent 2. \textbf{``Normalize'' hop-based link strengths}

Since 1-, 2-, and 3-hop results each have their own statistics, they must first be normalized to create an aggregate score.  To normalize the hop-based link strengths, first, the average link strength $\mu_n$ for each hop count $n$ is computed:  $\mu_n = \frac{1}{|E|} \sum_{e_{ij} \in E}[a_{n,ij}]$.  The average link strength is then used to calculate a normalized link cohesion $c_{n,ij}$ for all edges $e_{ij}$ with associated $n$ hop connections for $n \in \{1, 2, 3\}$:
\begin{equation*}
c_{n, ij} = \frac{a_{n,ij}}{\mu_n + a_{n,ij}}
\end{equation*}

This formulation scales the hop-based edge strengths to the interval $[0,1)$.  

\noindent 3. \textbf{Calculate link cohesion as a cumulative edge strength}

The scaled link strengths for the three $n$-hop connections are averaged to compute the aggregate scaled local edge strength, which we will call the link cohesion $c_{ij}$:
\begin{equation*}
c_{ij} =\frac{1}{3} [ c_{1, ij} + c_{2, ij}  + c_{3, ij} ]
\end{equation*}

In the calculation of link cohesion, the three normalized hop-based link strengths are weighted equally.  We investigated inclusion of the three terms as well as individual hop-based link strength contributions by assessing the associated performance of our edge-pruning community detection algorithm, defined in the next section, on an open-source real-world dataset, the European Union (EU) email data.  The performance results for various binary weightings are provided in Table~\ref{tab:coefficients}.  Performance was measured using the remaining number of edges, the number of communities detected, and \emph{F-score}, that is the harmonic mean of precision and recall to assess community detection performance as discussed in\cite{satuluri2011local}.  Each of the three path lengths provide value toward the link cohesion score.  While some combinations yielded detection of more communities or higher F-scores than others, there was not enough of a difference to justify using a formula other than equal weighting or to justify removal of any term.  This example demonstrates that inclusion of all three hop-based link strengths provides one of the highest F-scores with the most remaining edges intact by the pruning algorithm.  While only the first two hops could be leveraged for faster performance, the third hop can be efficiently calculated using 2-hop results.  
\vspace{-1em}
\begin{table}[!htb]
\caption{Example evaluation of relative hop contributions on the EU email data assessed using the remaining number of edges, number of communities detected, and F-score with the pruning algorithm defined in the next section.} \label{tab:coefficients}
\centering
\begin{tabular}{ c | c | c | c | c | c}
\multicolumn{3}{c|}{Binary inclusion of} &Remaining& Communities &\\
$ c_{1, ij}$ & $c_{2, ij} $ & $c_{3, ij}$  & Edges & Detected & F-score\\
 \hline 
 1 & 1 & 1 & 2801 & 17 of 42 & 0.539\\
 1 & 1 & 0 & 2118 & 17 of 42 & 0.530\\
 1 & 0 & 1 & 2189 & 15 of 42 & 0.517\\
 0 & 1 & 1 & 2120 & 18 of 42 & 0.517\\
 1 & 0 & 0 & 1725 & 13 of 42 & 0.286\\
 0 & 1 & 0 & 1645 & 22 of 42 & 0.469\\
 0 & 0 & 1 & 1990 & 16 of 42 & 0.544
\end{tabular} \\
\end{table}
\vspace{-1em}





\section{Using Link Cohesion}

Link cohesion metrics have multiple applications as described in this section.  We specifically explore how link cohesion can be used to simplify graphs through a pruning algorithm.

\subsection{Link Cohesion Density \& Pruning Algorithm}

Density as a concept is commonly used in a broad array of statistical learning algorithms, such as DBSCAN, Gaussian mixture models, and Galileo \cite{savkli2017galileo}.  In a similar way, we define a density concept here which we will use to support graph pruning through maximization of graph density.

The motivation for the \emph{Maximum Density Core} pruning algorithm, MDCore, is two-fold:  increase the average link cohesion weight in the graph while maximizing the number of connected vertices.  

In order to accomplish this, we define a metric \emph{link cohesion density} that accounts for both: 
\begin{equation}
\rho = v_c  c_{average}
\end{equation}
where $v_c$ is the number of vertices with degree greater than 0 and $c_{average}$ is the average cohesion score for the remaining edges.  While our focus in this paper is a global density metric, this concept of density could be used to characterize local density as well, for example, to generate a heatmap view of the graph.

In MDCore, edges are deleted, starting with the weakest cohesion score, until $\rho$ reaches a maximum density.  Link cohesion scores are not recalculated during the pruning process since they are informed by the original graph; this also reduces computation time.
Because weak edges are associated with high degree vertices on both ends, their removal increases the average cohesion score without initially decreasing the number of vertices.  Continued link deletions eventually lead to the optimal maximal average, after which it decreases as the graph becomes more and more disconnected.   

\begin{figure}[!htb]
\centering
	\begin{subfigure}[b]{0.45\textwidth}
		\includegraphics[width = \linewidth]{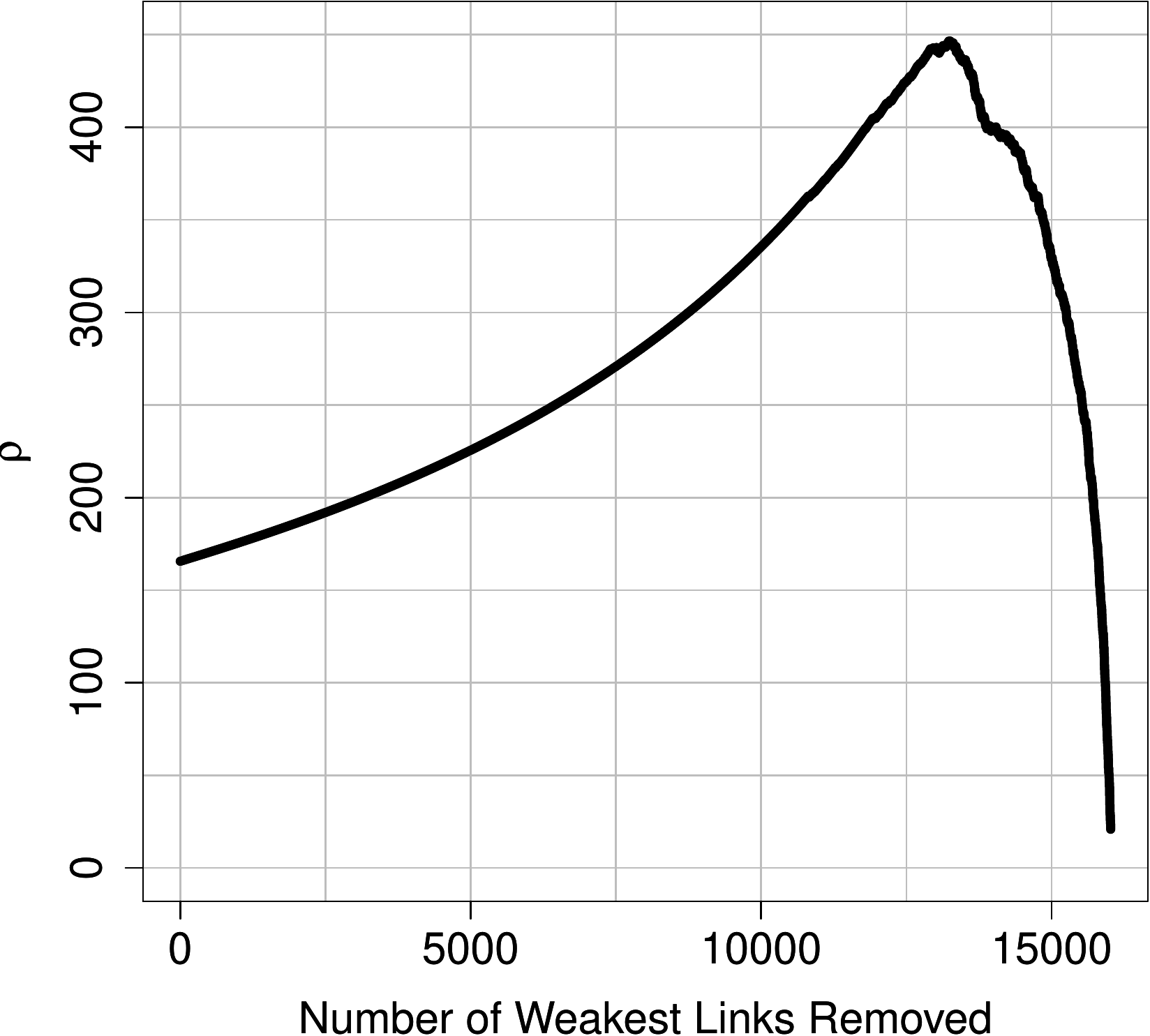}
		\caption{}
	\end{subfigure}
	\begin{subfigure}[b]{0.45\textwidth}
		\includegraphics[width = \linewidth]{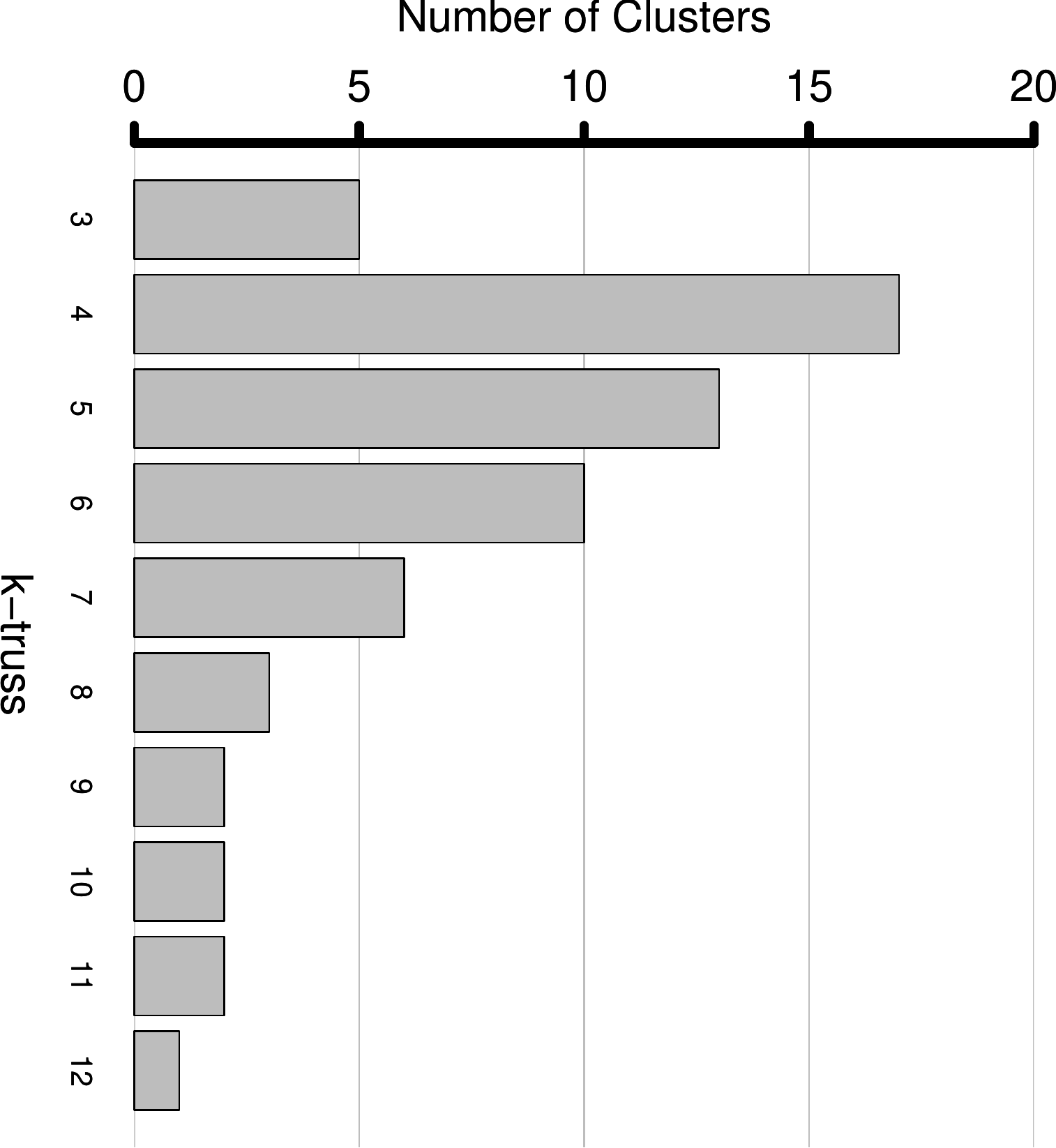}
		\caption{}
	\end{subfigure}
\caption{MDCore pruning algorithm applied to the EU email network:  (a) graph density $\rho$ as a function of the number of weakest links removed and (b) number of clusters found using various truss sizes; only one cluster could be found on this network when links were not removed whereas a maximum of 17 clusters were found at truss level 4 after pruning.}
\label{fig:eu_analysis}
\end{figure}

This behavior can be seen in the plot of $\rho$ values for the EU email network in Figure~\ref{fig:eu_analysis}(a).  This plot is used to identify a link cohesion score threshold, corresponding to the number of weakest links removed, below which all other links will be removed.  In this example, we find that out of the original $\sim$16k links, removal of $\sim$12k links provides the maximum density graph.

In order to demonstrate the benefit of removing edges, we compare truss-finding on the base graph to truss-finding on the pruned graph,  where a $k$-truss is a sub-graph where each edge in the subgraph is supported by $k-2$ triangles.  The truss-finding algorithm we employ identifies trusses at different levels \cite{cohen2008trusses, cohen2009graph, cohen2019trapeze} and selects the truss level where the number of clusters is maximized.  At the lowest level, most of the graph is part of the truss while at the other extreme a minority of nodes are left in the truss.  In between, there is an area where the number of truss-based communities are maximized.  Whenever the number of clusters are equal between levels, the lower level is chosen to include a larger segment of the graph.  We refer to this algorithm as maximal-community truss-finding.  We specifically use truss-finding as an example of how this edge removal method can help expedite community detection analysis.  The relationship between the number of nodes in the cluster and the number of clusters can be seen in Figure~\ref{fig:eu_analysis}(b), an example applying a combination of MDCore and the maximal-community truss-finding algorithm to the EU Email network dataset \cite{paranjape2017motifs}; in this case the number of communities is maximized to 17 at truss level 4.  When maximal-community truss-finding was applied to the original graph, without pruning, only one cluster was found at every truss-level and no sub-communities could be detected.

\subsection{Other Potential Uses of Link Cohesion}

Link cohesion is specifically useful for highly connected networks such as cyber and social networks, where it can be used to pre-process and simplify datasets.  Link cohesion is unlikely to be useful for physical networks such as infrastructure with low average degree.  In considering available networks in the SNAP database \cite{snapnets}, nearly 40\% of datasets have an average node degree greater than 10.  We will specifically explore performance for comparable highly connected conditions.

Link cohesion may also serve as a fast, inexpensive approximation of edge betweenness since link cohesion accounts for the extent to which a degree is central at least locally.  We explore this correlation in Section~\ref{sec:corr_results}.

Link cohesion may also be used as edge weights to support graph analysis algorithms such as community detection.   This application is not explored in this paper, beyond the maximal-community truss-finding algorithm.




\section{Relevant Literature} \label{sec:lit_rev}

While multiple techniques have been developed to incorporate edge weights (e.g. \cite{newman2004analysis}), very few edge metrics exist.   One popular nodal metric, betweenness centrality, a measure based on the number of shortest paths that flow through a node, was generalized to edges in 2002 by Girvan and Newman \cite{girvan2002community}.  Newman subsequently used another edge metric to support calculating the modularity of communities \cite{newman2006modularity}; he also specifically leveraged the expected number of edges between nodes.  De Meo et al. developed the $\kappa$-path edge centrality metric \cite{de2012novel} which uses fixed length random walks to efficiently compute the importance of edges.  Harel et al. developed a dissimilarity-based edge metric to model visual saliency \cite{harel2007graph}.

Edge removal techniques have been used in a variety of algorithms to help simplify networks.
For example, \emph{edge filtering} (see \cite{glattfelder2013backbone} and \cite{tumminello2005tool}), uses edge weights to remove edges to reduce graphs to a set of critical edges or a network backbone.  These techniques rely on edge weights and do not support statistical graph analysis since too many edges are removed.

As another example, \emph{similarity sparsification} has been employed to intentionally remove edges based on similarity of adjacent neighbors (i.e. considering all 2-hop connections) local and global thresholds \cite{satuluri2011local}.  Similarity between nodes $i$ and $j$ for edge $e_{ij}$ is calculated as $Sim(i,j) = |Adj(i) \cap Adj(j) | / |Adj(i) \cup Adj(j) |$
where $Adj(i)$ is the set of adjacent neighbors to node $i$. 
Satuluri et al. found that sparsification of edges can enable both higher clustering accuracy and significant speedups (10-50 fold).  As such, we will leverage the local \emph{Sparsify} algorithm \cite{satuluri2011local} as a performance benchmark.

Numerous community detection algorithms directly leverage edge removals.  
The Girvan-Newman algorithm uses progressive edge removals, selected for high edge betweenness, in order to perform community detection \cite{girvan2002community}.  
Bai et al.~recently introduced a community detection algorithm which uses networks simplified through identification of nodes' leading and following degrees \cite{bai2018novel}.  
Modularity, a concept that enables comparison of the actual and expected number of edges between two communities, has also been used for community detection \cite{newman2006modularity}.  
The Louvain algorithm expands on this concept using a heuristic to optimize modularity \cite{blondel2008fast}.




\section{Results} \label{sec:results}

\subsection{Experimental Datasets}

To evaluate the utility of the link cohesion algorithm, we will consider both real world and generated networks.  Since our demonstration employs a community detection algorithm, we focus on networks with non-overlapping communities.

Three real world network datasets have been selected, as listed in Table~\ref{tab:data}.  Zachary's Karate Club \cite{zachary1977information} and the Students (community 4 only) \cite{moody2001peer} networks provide relatively small baseline networks.  The Karate Club network has two underlying communities.  
The Students network has two schools with students in six different grades, with some students not in a specified grade, as well as gender and ethnicity attributes. As such, the number of ground truth communities (*) is unclear for the Students network.
One can study how to objectively define communities for networks with multiple features in real-world networks; however, this was not considered in this paper.  
The European (EU) email network \cite{paranjape2017motifs} provides a highly connected and thereby challenging example with its edge-to-node ratio of 25.  
Datasets are available through \cite{freemandata} and \cite{snapnets}.
\vspace{-1em}
\begin{table}[htb]
\caption{Example Real-world Networks} \label{tab:data}
\centering
\begin{tabular}{ l | c | c | c }
 Dataset & Nodes & Edges & Communities \\
 \hline 
 Karate Club \cite{zachary1977information} & 34 & 78 & 2 \\ 
 Students Comm 4 \cite{moody2001peer} & 291 & 1396 & 2 or 7*  \\
 EU email core \cite{paranjape2017motifs}  & 1005  & 25k & 42 
\end{tabular} \\
\end{table}

We demonstrate the use of link cohesion on Lancichinetti-Fortunato-Radicchi (LFR) generated networks as well \cite{lancichinetti2008benchmark}.  LFR networks simulate real-world networks with variable community sizes and degrees.  To generate LFR networks, users can specify the number of nodes $N$, the average degree $\langle k 
\rangle$, the maximum degree $k_{max}$, a degree distribution scaling parameter $t_1$, a community size distribution scaling parameter $t_2$, and the mixing parameter $\mu$ in order to perform controlled evaluations of algorithms.  The mixing parameter $\mu$ describes the average fraction of out-group connections and will be varied to study performance under several mixing conditions.  We will also examine how performance scales with increasing graph size by varying $N$.

\subsection{Edge Pruning Study}

To demonstrate the use of edge removals, we first consider how our algorithm performs when using the maximum density threshold to prune both real and generated networks before applying a clustering algorithm.  In these examples, we employ our maximal-community truss-finding algorithm for clustering. 

To evaluate the performance of edge pruning on real-world and LFR networks, two metrics will be used:  number of detected communities (DC) and F-score.  We will also consider processing time for analysis of the LFR networks.

\begin{wraptable}{r}{0.57\textwidth}
\caption{Results of maximal-community truss-finding with various pruning algorithms on real-world networks} \label{tab:real_results}
\centering
\begin{tabular}{ l | c | c | c }
 Dataset & Pruning Algorithm & DC & F-score
	\\
 \hline 
\multirow{3}{*}{Karate Club } 
	 	& Unpruned & 2 & 0.52 \\
	 	& Sparsify \cite{satuluri2011local} & 1 & 0.48\\
	 	& MDCore & 0 & --\\
  		\hline 
\multirow{3}{*}{Students } 
	 	& Unpruned & 7 & 0.42 \\
	 	& Sparsify & 10 & 0.17 \\
	 	& MDCore & 13 & 0.24 \\
  		\hline 
\multirow{3}{*}{EU email}
	 	& Unpruned & 1 & 0.19 \\
	 	& Sparsify & 17 & 0.28 \\
	 	& MDCore & 17 & 0.54\\
\end{tabular} \\
\end{wraptable}

Results on the three real data sets are provided in Table~\ref{tab:real_results}.  On the smaller graphs, the MDCore algorithm has the lowest F-score, indicating poorer performance; however, on the more highly connected EU dataset, MDCore has the highest F-score, showing better performance.  

\begin{figure*}[!htb]
	\begin{subfigure}[b]{0.3\textwidth}
		\includegraphics[height=2.5cm]{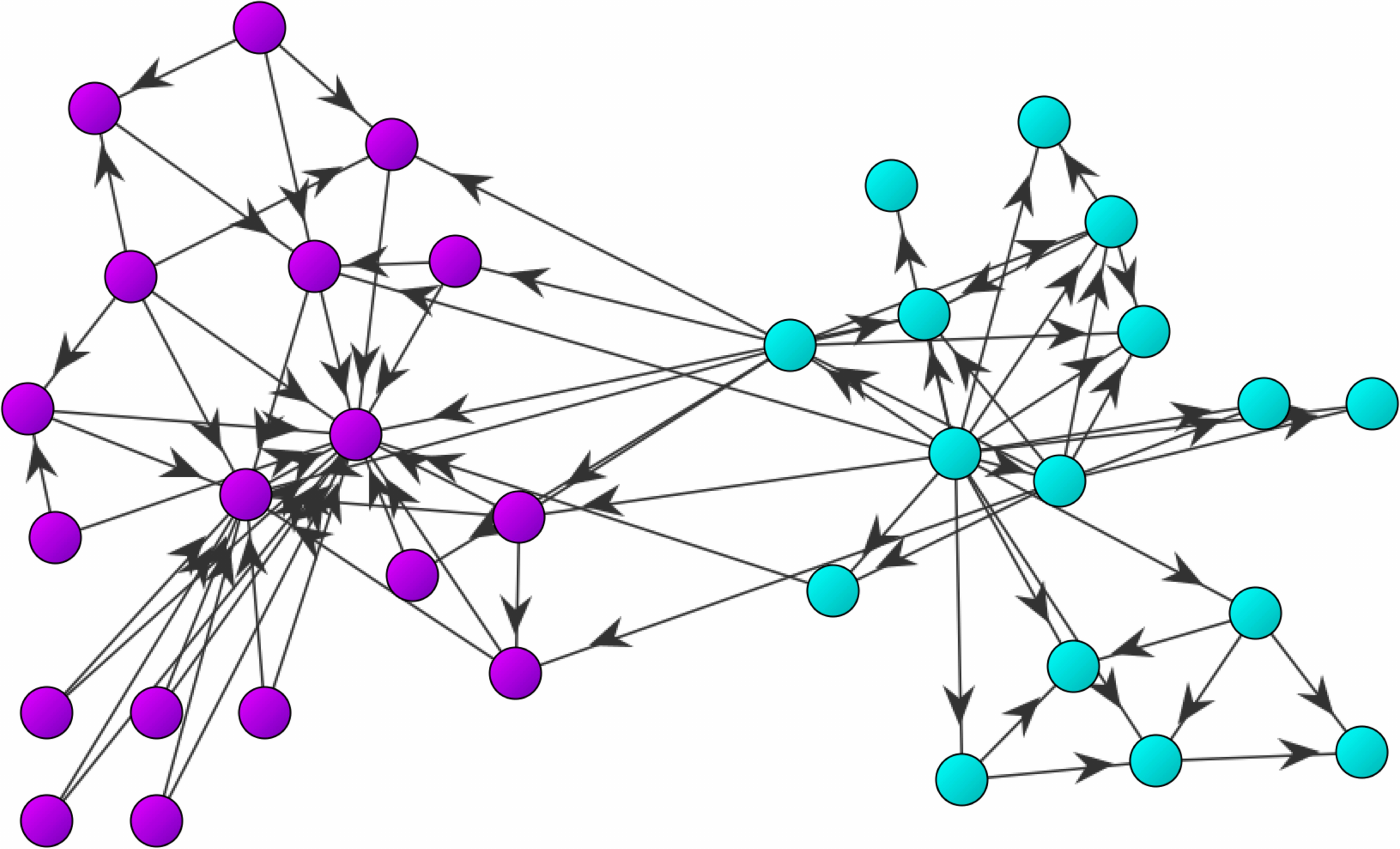}
		\caption{}
	\end{subfigure}
	\begin{subfigure}[b]{0.3\textwidth}
		\includegraphics[height=2.5cm]{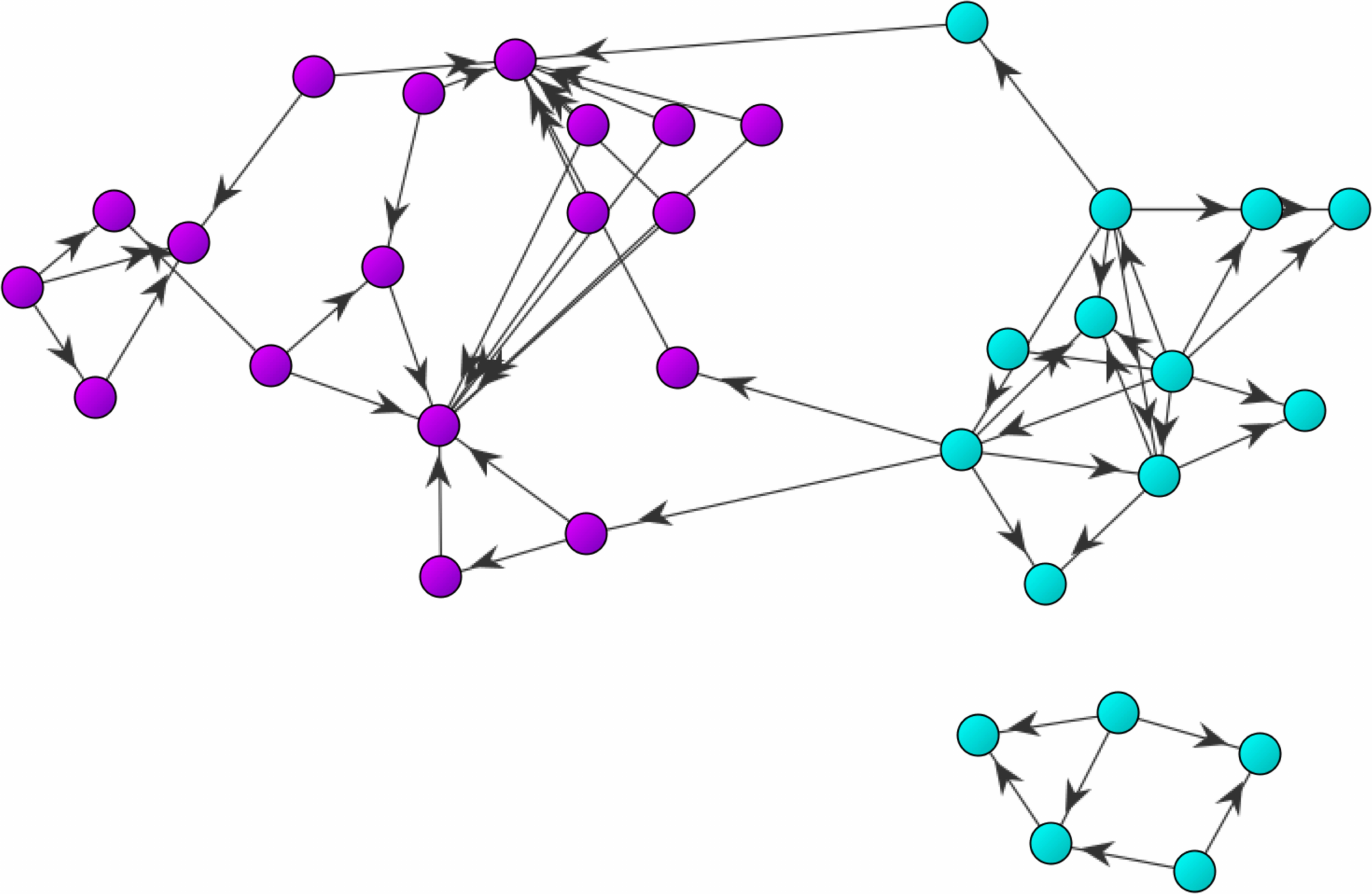}
		\caption{}
	\end{subfigure}
	\begin{subfigure}[b]{0.3\textwidth}
		\includegraphics[height=2.5cm]{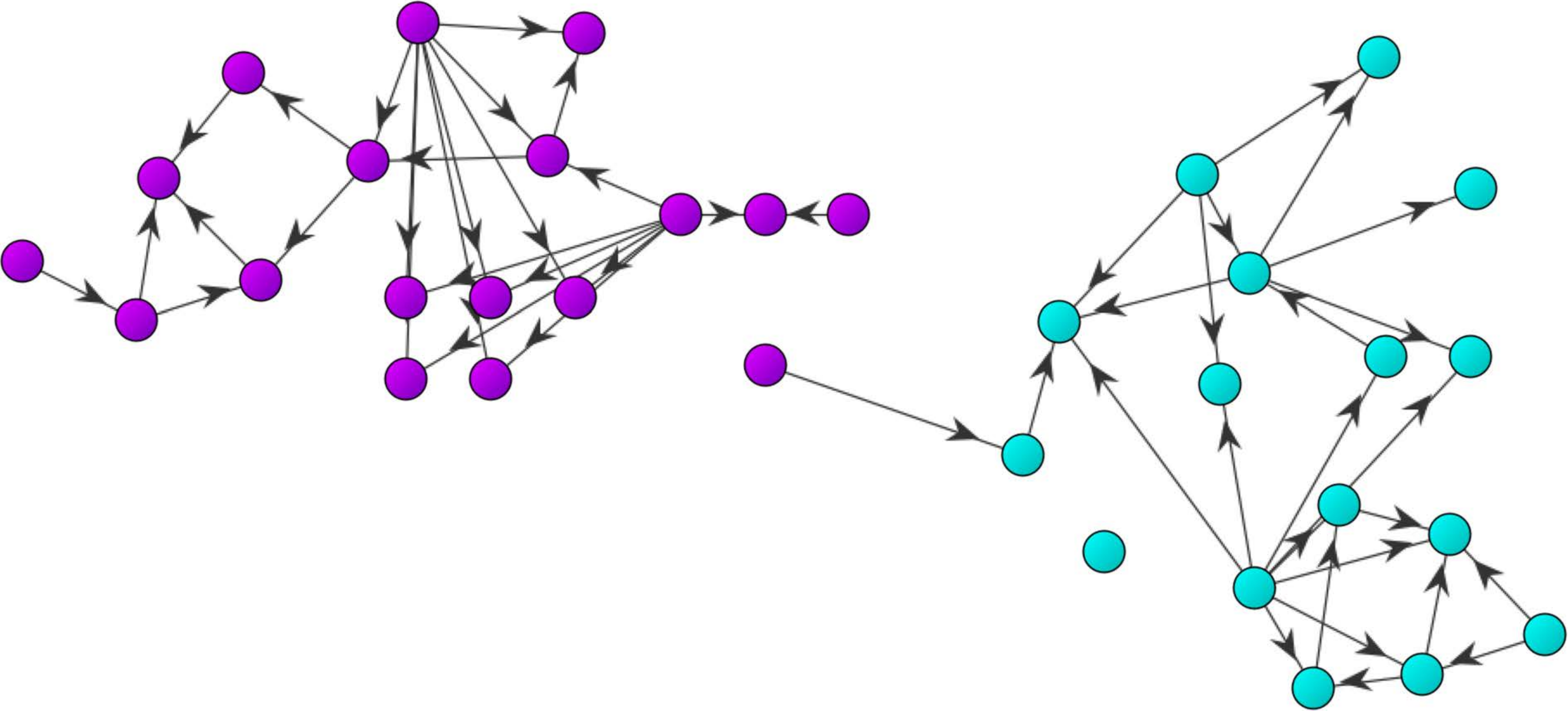}
		\caption{}
	\end{subfigure}
\caption{The Karate dataset with various edge removals techniques applied: (a) original edge set, (b) Sparsify edge set, and (c) MDCore edge set}  \label{fig:karate}
\end{figure*}

\begin{figure*}[!htb]
	\begin{subfigure}[b]{0.3\textwidth}
		\includegraphics[height=2.5cm]{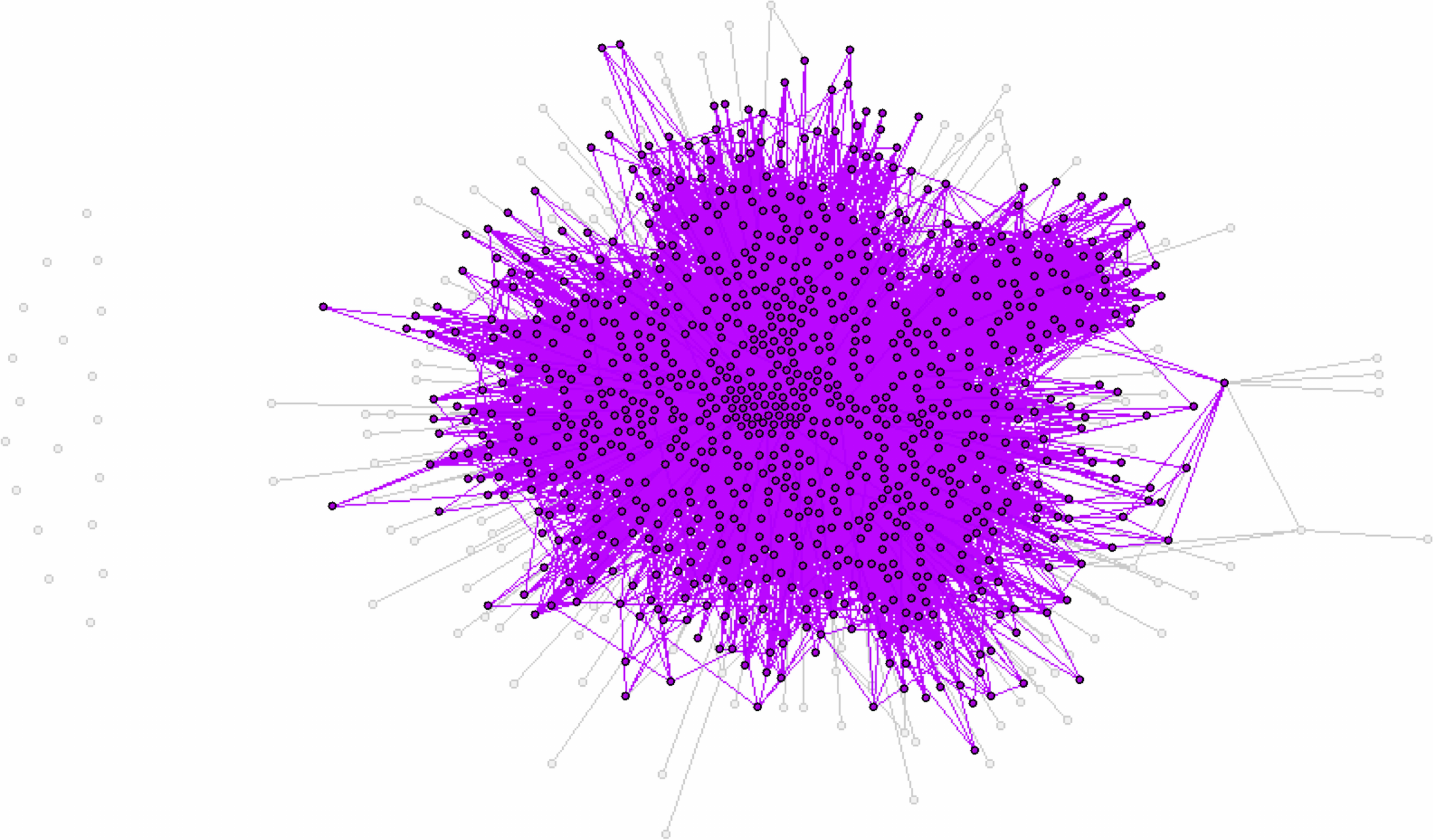}
		\caption{}
	\end{subfigure}
	\begin{subfigure}[b]{0.3\textwidth}
		\includegraphics[height=2.5cm]{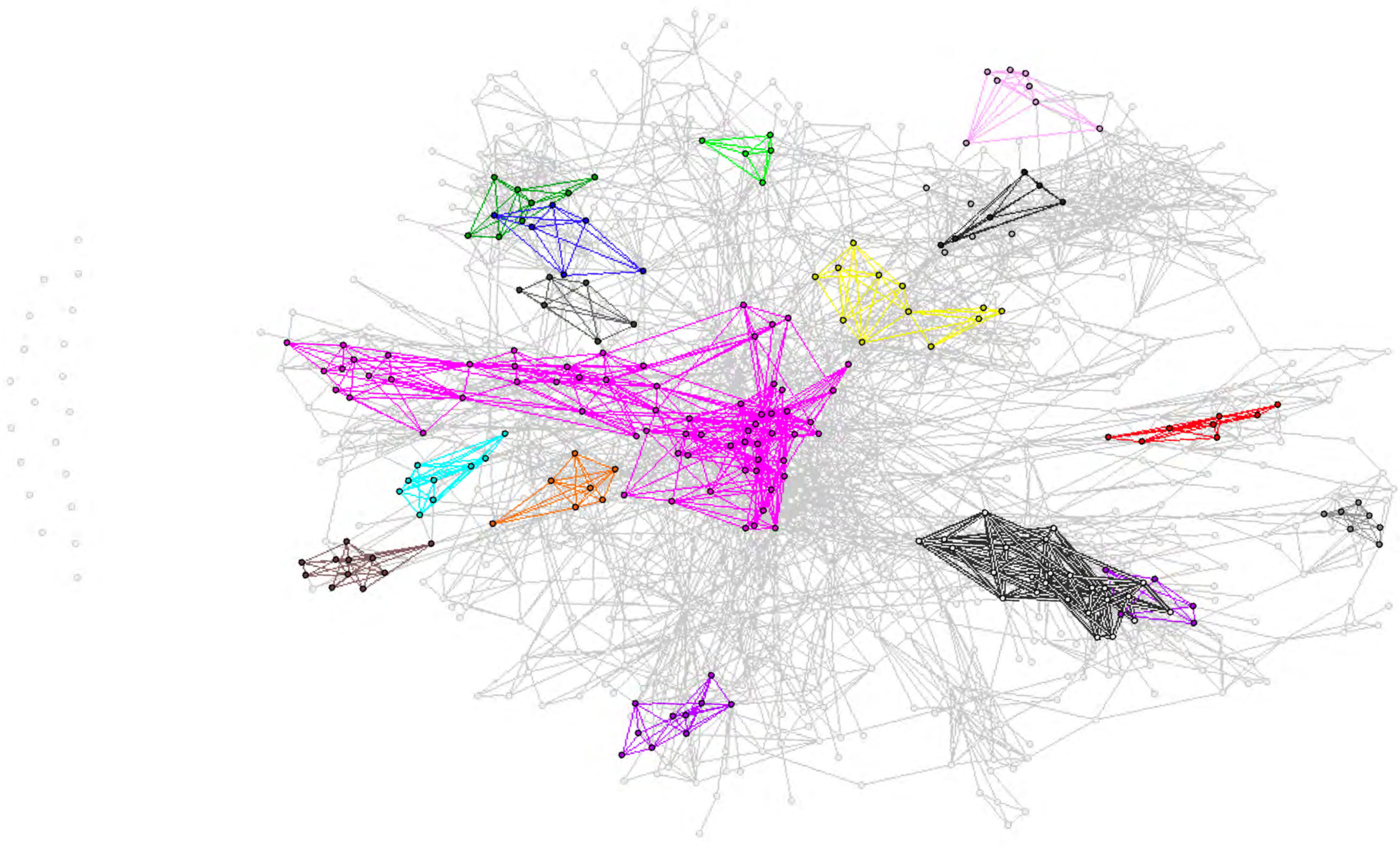}
		\caption{}
	\end{subfigure}
	\begin{subfigure}[b]{0.3\textwidth}
		\includegraphics[height=2.5cm]{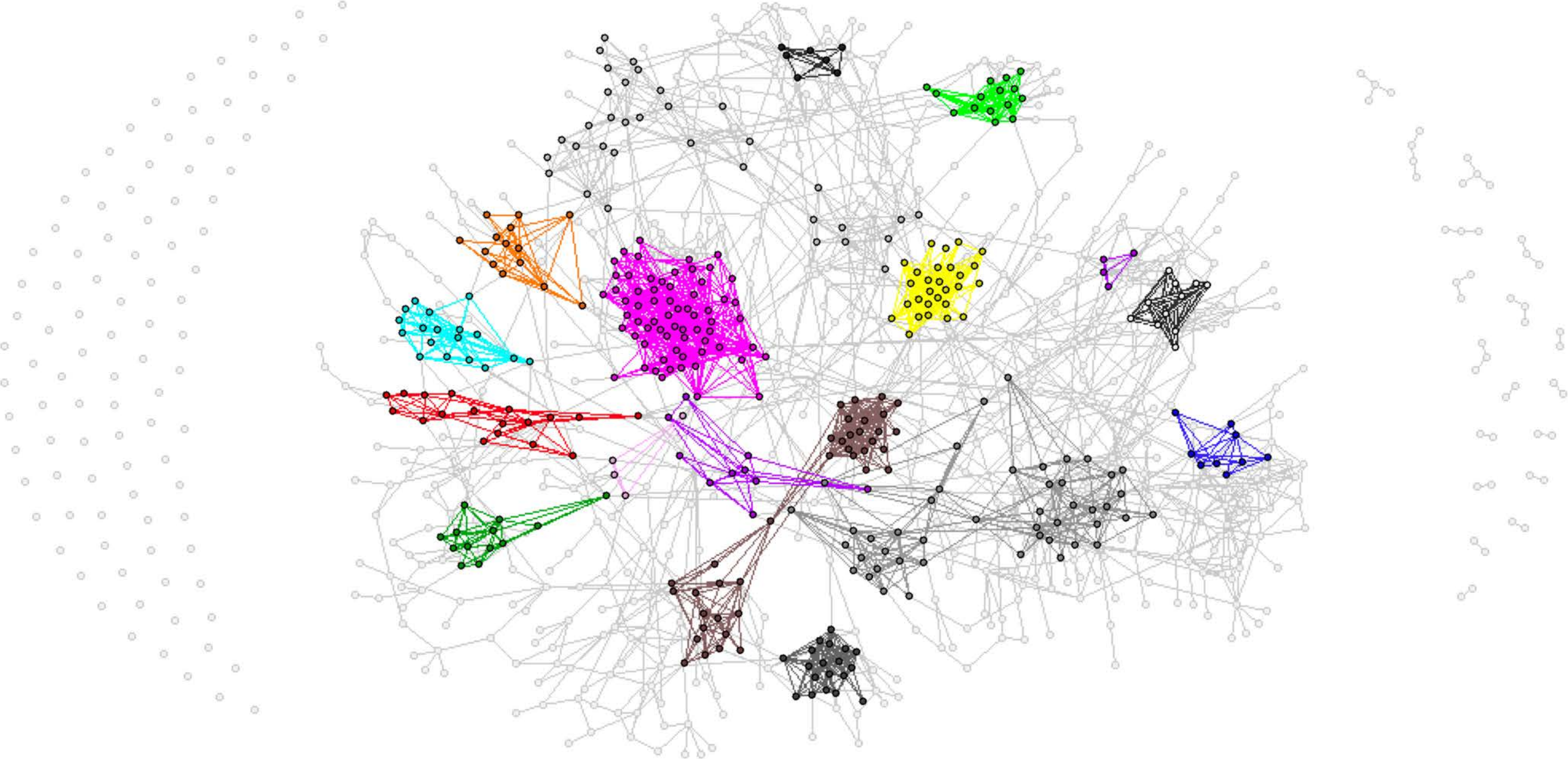}
		\caption{}
	\end{subfigure}
\caption{Clusters identified through maximal-community truss-finding on the EU email dataset using (a) original edge set, (b) Sparsify edge set, and (c) MDCore edge set} \label{fig:euemail}
\end{figure*}

While the F-score could not be calculated when MDCore was applied to the Karate Club dataset, the graph divides most cleanly with MDCore, as shown in Figure~\ref{fig:karate}, with colors representing ground truth community membership.  On the EU email dataset, as shown in Figure~\ref{fig:euemail}, we observe that link removals support the identification of communities which could not be readily discerned by maximal-community truss-finding on the original dataset; here different colors are used to show the detected community membership.  
This analysis of real world datasets shows that MDCore may provide an advantage when applied to highly connected datasets, which we will further examine using generated LFR networks.  

\begin{figure*}
\begin{subfigure}{\linewidth}
	\begin{subfigure}[t]{0.24\textwidth}

		\includegraphics[width =\linewidth]{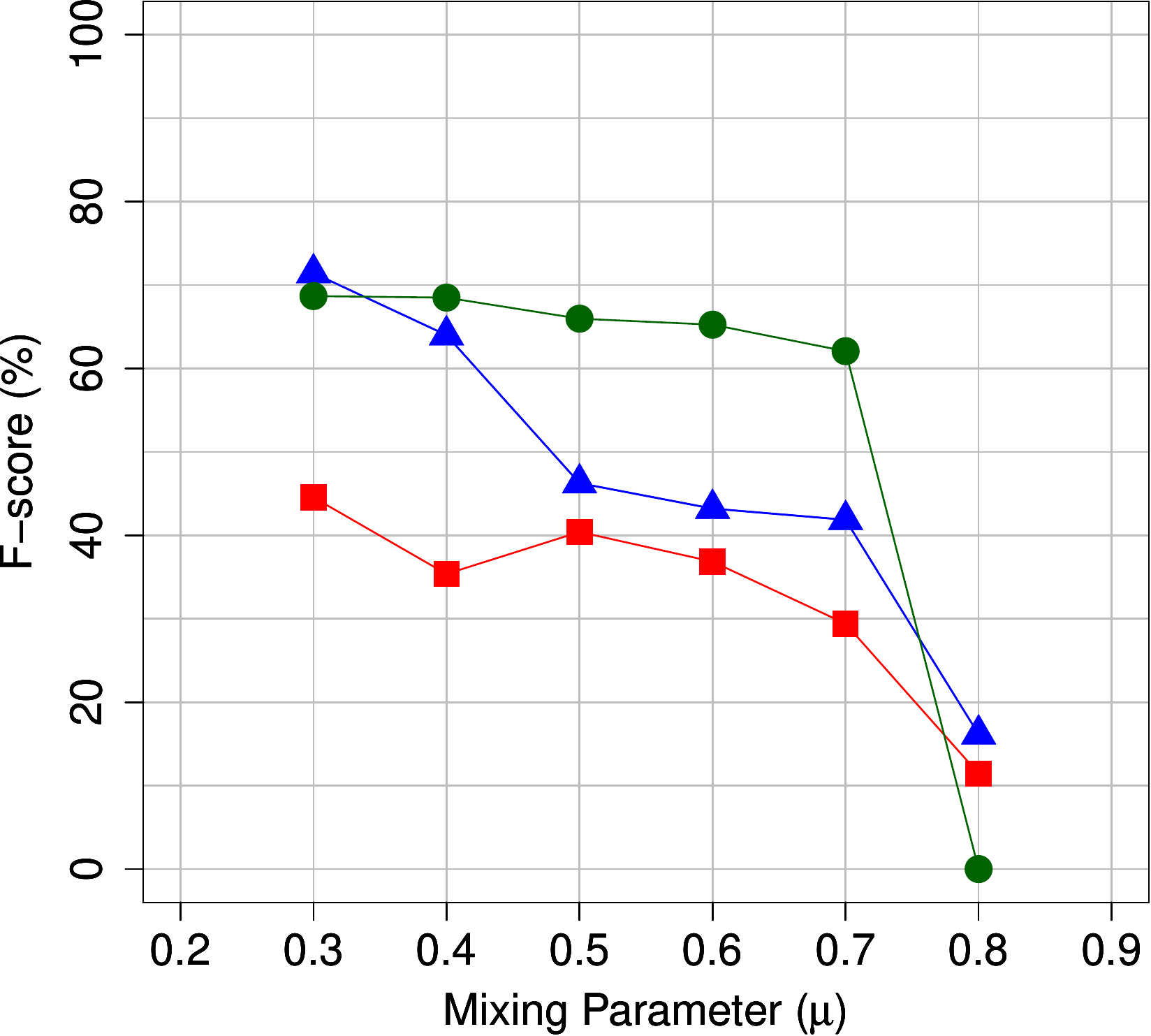}
	\end{subfigure}
	\begin{subfigure}[t]{0.24\textwidth}
		
		\includegraphics[width =\linewidth]{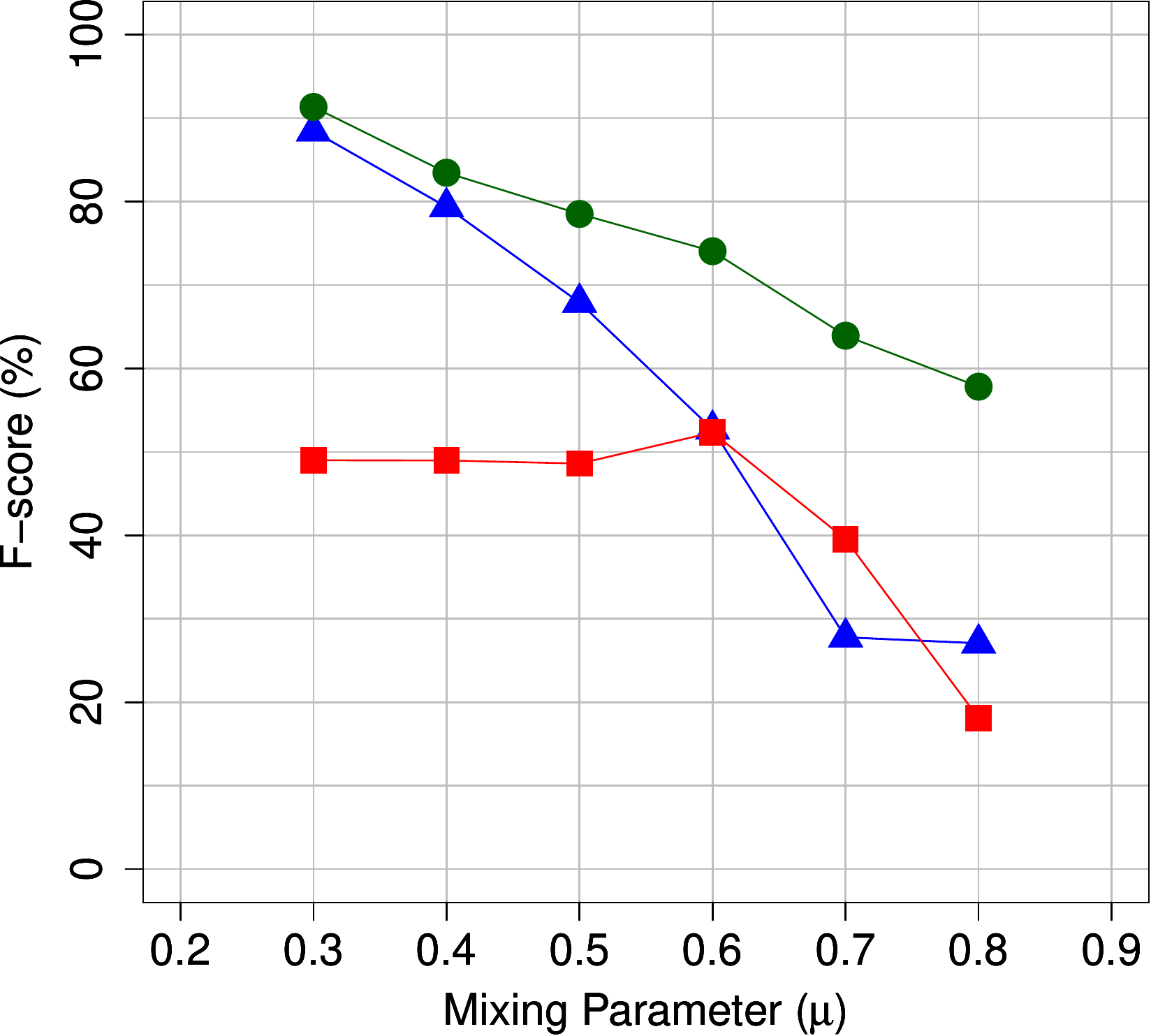}
	\end{subfigure}
	\begin{subfigure}[t]{0.24\textwidth}
		
		\includegraphics[width =\linewidth]{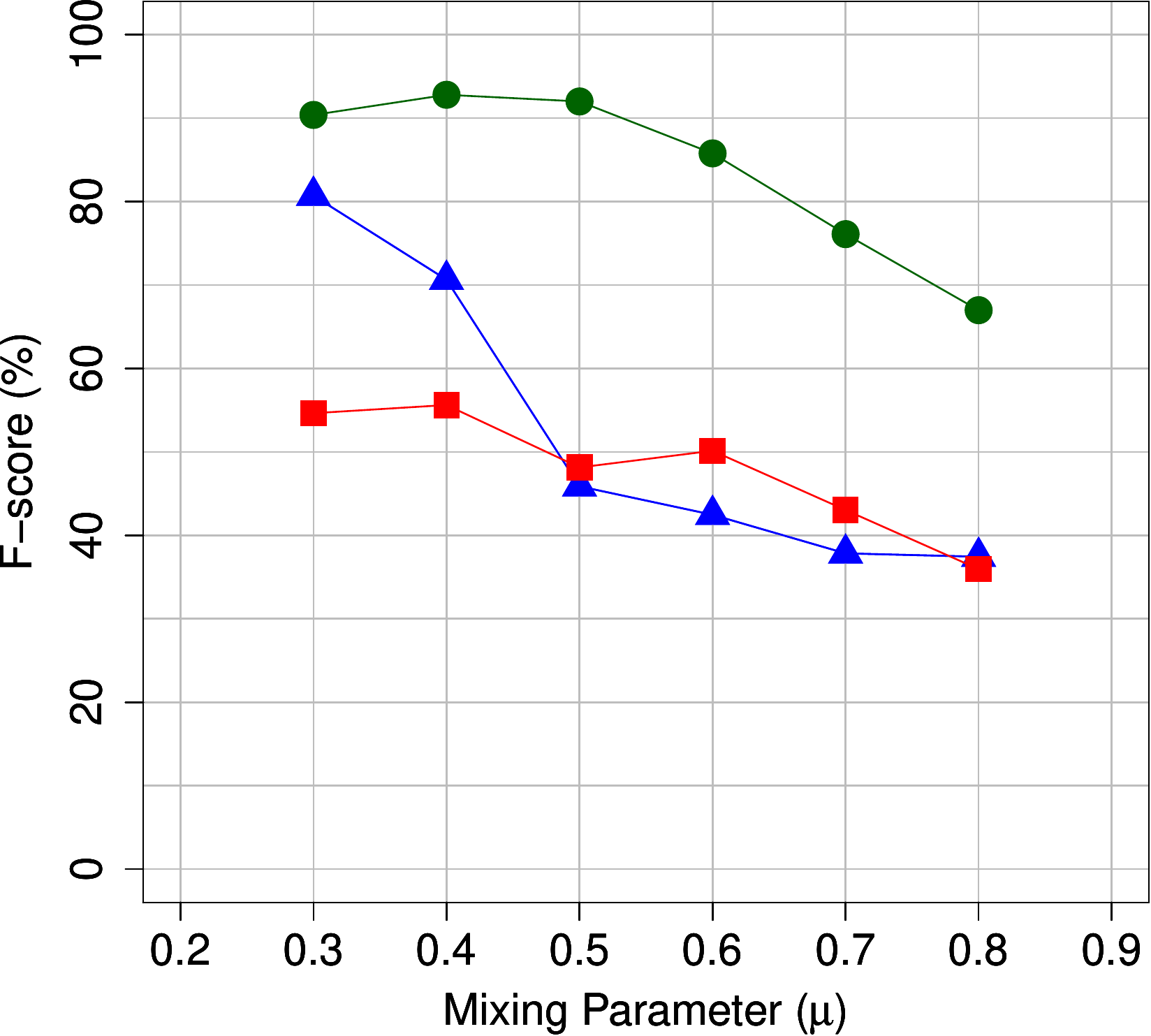}
	\end{subfigure}
	\begin{subfigure}[t]{0.24\textwidth}
		\includegraphics[width =\linewidth]{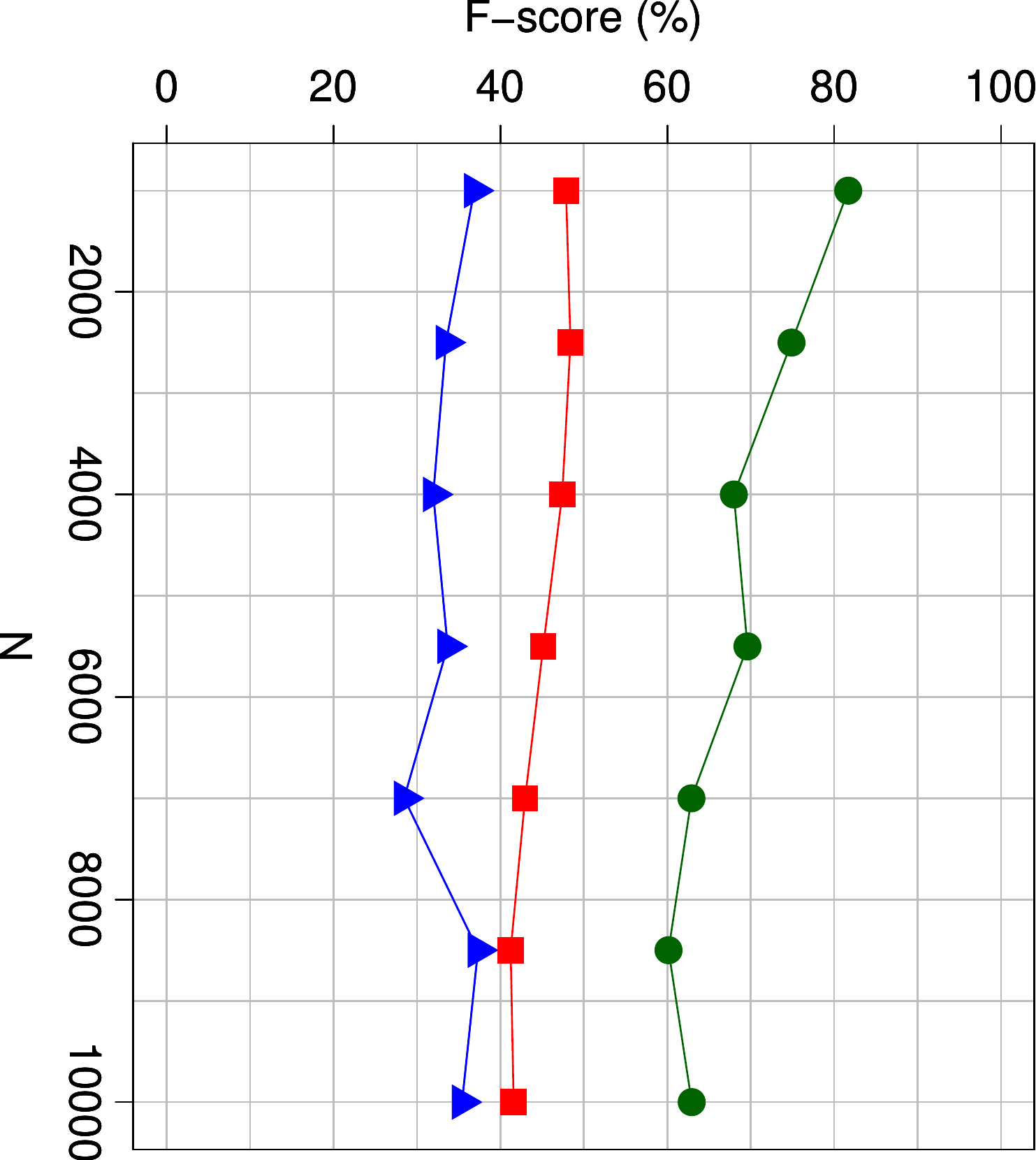}
	\end{subfigure}
\end{subfigure}
\begin{subfigure}{\linewidth}
	\begin{subfigure}[b]{0.24\textwidth}
		\includegraphics[width =\linewidth]{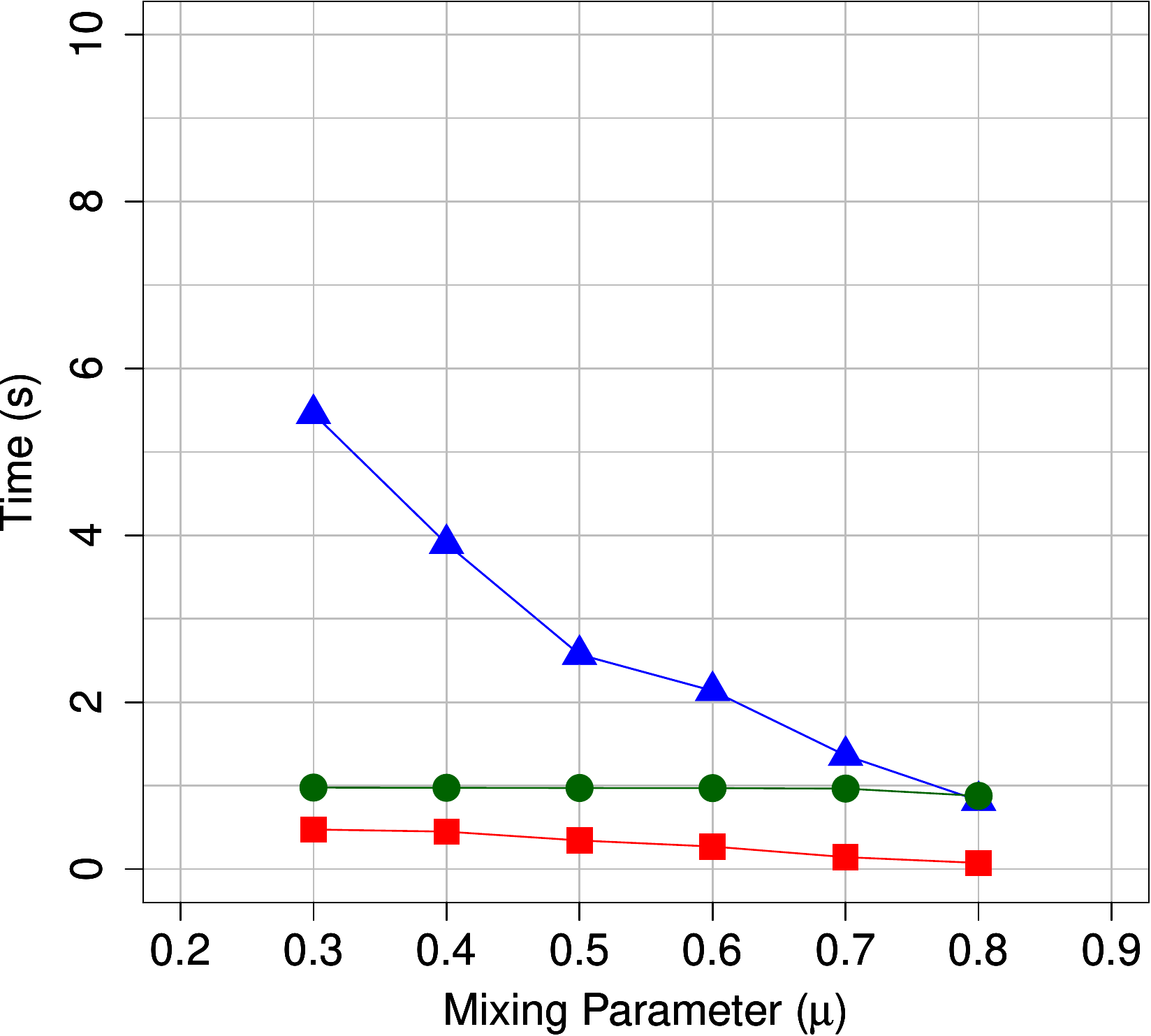}
		\caption{$k$=10, $k_{max}$=100}
	\end{subfigure}
	\begin{subfigure}[b]{0.24\textwidth}
		\includegraphics[width =\linewidth]{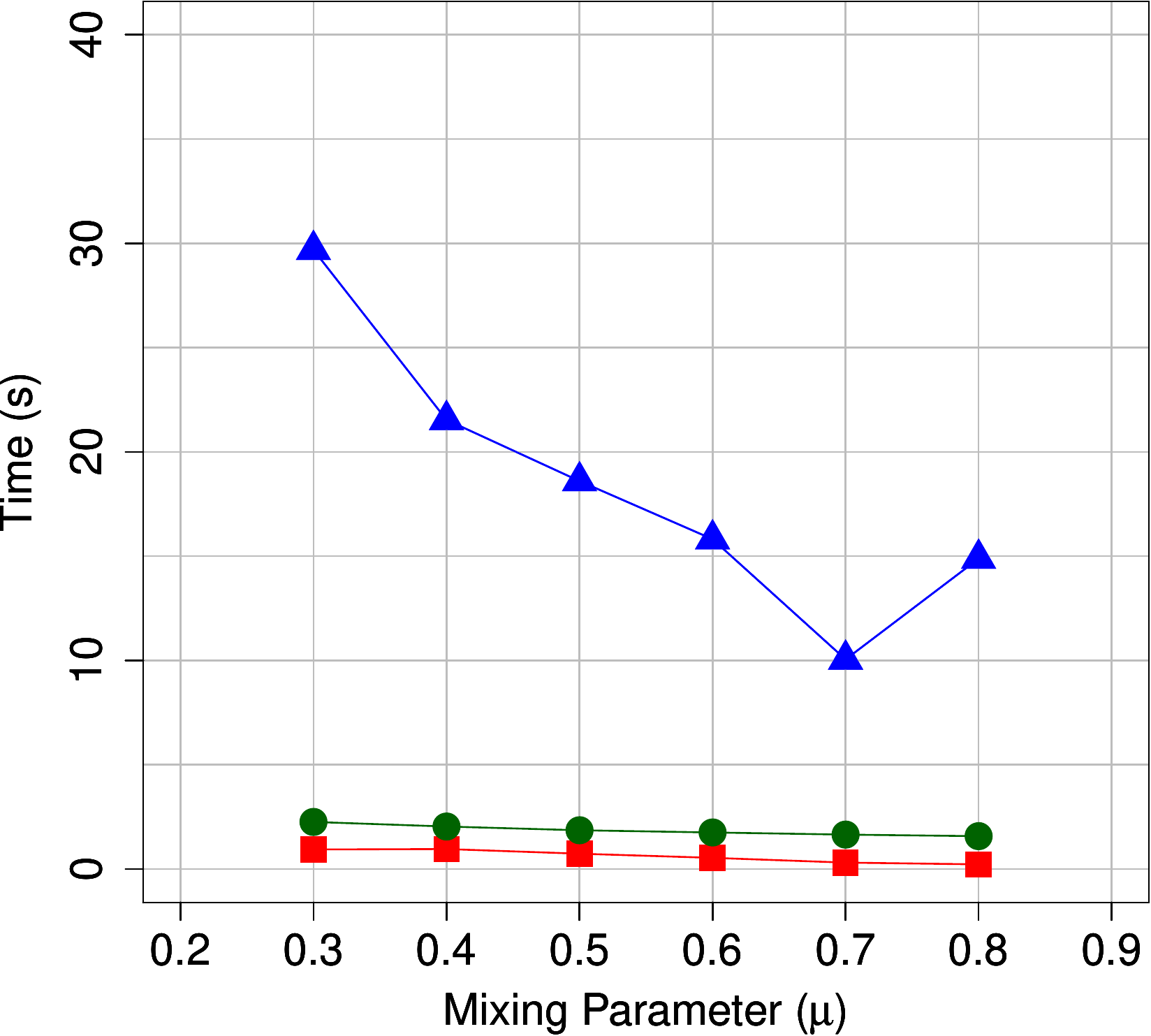}
		\caption{$k$=20, $k_{max}$=200}
	\end{subfigure}
	\begin{subfigure}[b]{0.24\textwidth}
		\includegraphics[width =\linewidth]{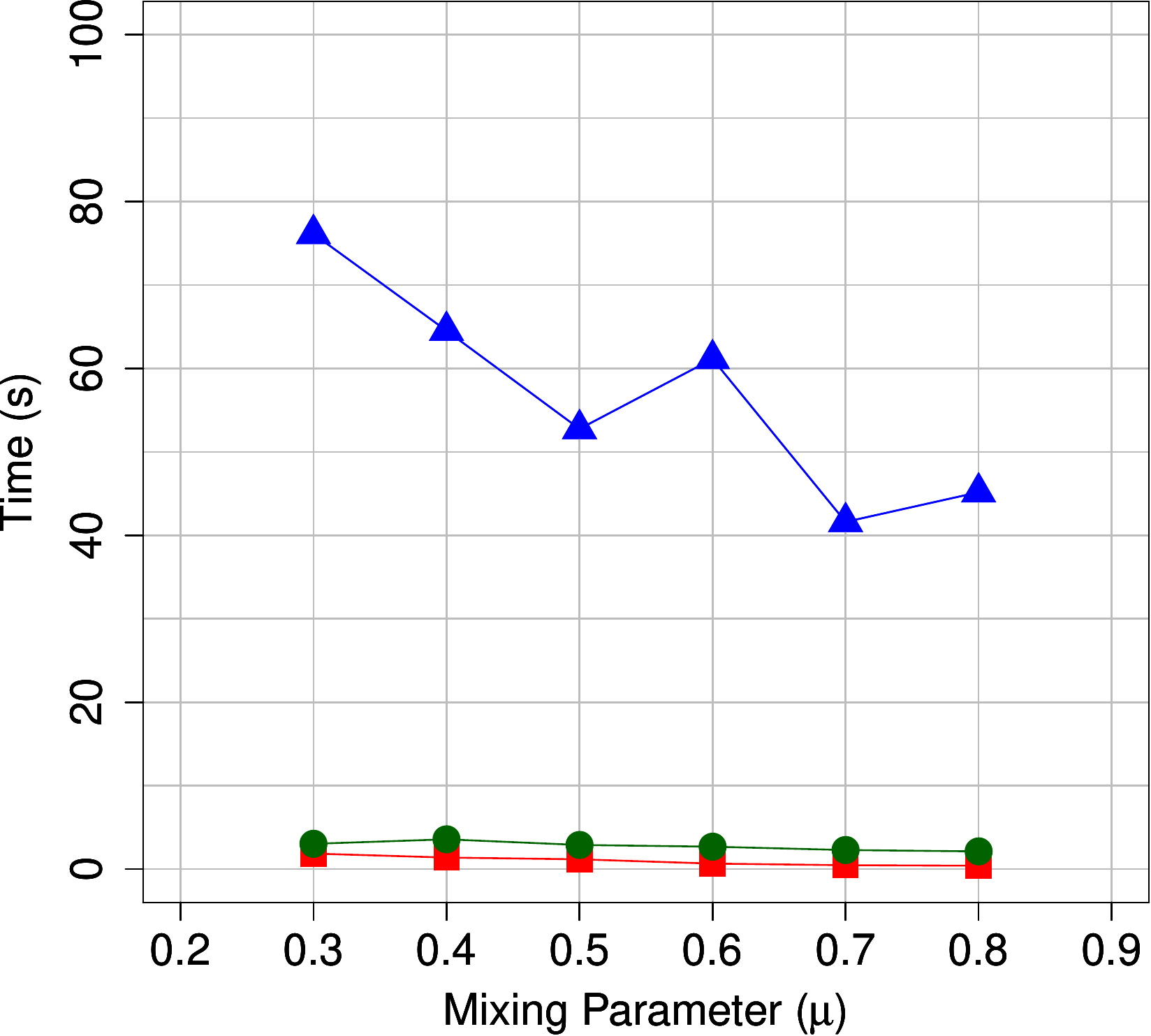}
		\caption{$k$=30, $k_{max}$=300}
	\end{subfigure}
	\begin{subfigure}[b]{0.24\textwidth}
		\includegraphics[width =\linewidth]{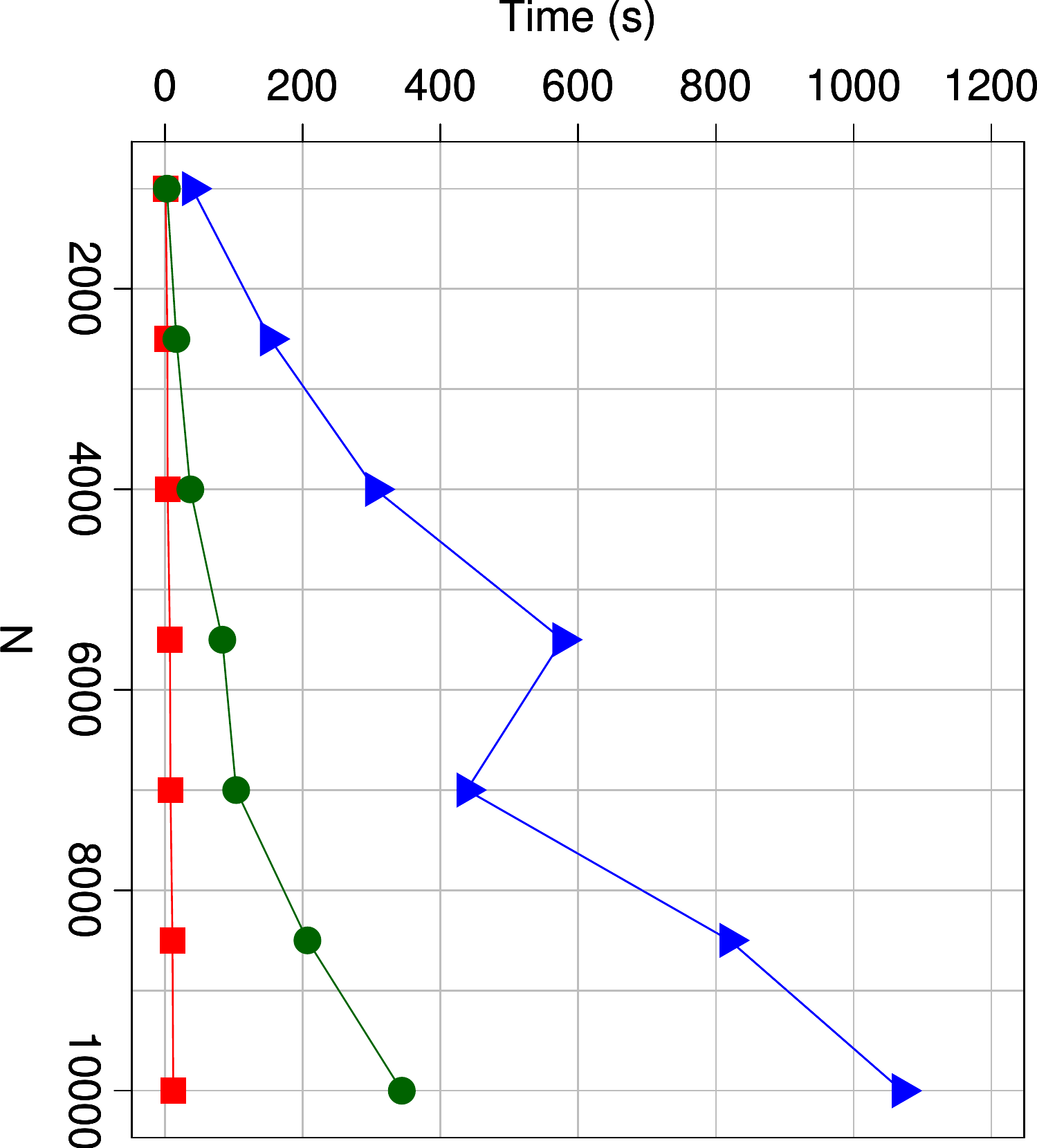}
		\caption{$k$=30, $k_{max}$=$\frac{N}{3}$}
	\end{subfigure}
\end{subfigure}

\caption{F-score and timing performance of maximal-community truss-finding algorithm on the original (blue triangles), Sparsify (red squares), and MDCore (green circles) networks for LFR networks generated with scaling parameters $t_1$ of 2.0 and $t_2$ of 1.5 with increasing average degree $k$ and maximum degree $k_{max}$ over a span of mixing parameters $\mu$.  In (a)--(c), $N$=1000.  In (d), $\mu$=0.6.  
}
\label{fig:lfr_inc_k_results}
\end{figure*}

The same algorithms are applied to randomly generated LFR networks with 1000 nodes, as shown in Figures~\ref{fig:lfr_inc_k_results}(a)--(c).  For this analysis, performance metrics are F-score and algorithm time.  For F-score, we observe that the performance of maximal-community truss-finding with the MDCore algorithm using link cohesion to identify removals outperforms both maximal-community truss-finding on the original graph and on the sparsified graph in nearly all cases.  This is particularly true for graphs with higher average degrees, in this case 30, as shown in Figure~\ref{fig:lfr_inc_k_results}(c).  
This is consistent with our working theory that link cohesion is most interesting and of greatest potential use for highly connected graphs with high average degrees.  
The Sparsify algorithm appears faster, although the MDCore algorithm is comparable relative to the time required to perform truss-finding on the original graph.  
Note that the Sparsify algorithm was written in C code whereas the original and MDCore were implemented in Java, so timing results are relative.

We further explore the impact of increasing the number of nodes to assess how algorithm performance scales.  Using aggregate results across five randomly generated LFR networks, the F-score and timing performance results are shown in Figure~\ref{fig:lfr_inc_k_results}(d) with corresponding tabulated results for the averaged number of communities detected in Table~\ref{tab:scaled_comm_detection}.  We observe that the MDCore algorithm has a consistently higher F-score for this parameter set.  Similar to the EU network analysis, the pruning of edges through density optimization leads to more communities and yields a higher F-score.  On this parameter set, we also observe that the Sparsify algorithm retains roughly 21\% of edges whereas the MDCore algorithm retains only about 5\% of edges after removals are performed.  

\begin{table}[!htb]
\caption{Community detection scaling study for truss-finding applied to randomly generated LFR networks average degree $k$ of 30, maximum degree $k_{max}$ of $N$/3, scaling parameters $t_1$ of 2.0 and $t_2$ of 1.5, and mixing parameter $\mu$ of 0.6. The corresponding F-score performance results are provided in Figure~\ref{fig:lfr_inc_k_results}(d).} \label{tab:scaled_comm_detection}
\centering
\begin{tabular}{ c | c | c | c | c }
& Ground & \multicolumn{3}{c}{Average DC for }\\
$N$ & Truth & Original & Sparsify & MDCore \\
 \hline 
1000 &	19.8 &	1.2 &	13.2 &	13.2\\
2500	& 35.8	& 1 &	22.2 &	26.4\\
4000 &	53.2 &	1 & 	25.4	&34.4\\
5500 &	50.6	& 1 & 	22.8	& 32.6\\
7000	 & 87.4	& 1 &	34.8	 & 42\\
8500 & 65.4	& 1.2	& 29	&34.6\\
10000 &	80.2 &	1	& 38	&43.4
\end{tabular} \\
\end{table}

\subsection{Edge Weights Study:  Correlations with Edge Betweenness} \label{sec:corr_results}

To further demonstrate the use of link cohesion, this section considers how link cohesion as edge weights relate to edge betweenness.  

\begin{figure*} [!htb]
\begin{subfigure}{\linewidth}
	\begin{subfigure}[t]{0.24\textwidth}

		\includegraphics[width =\linewidth]{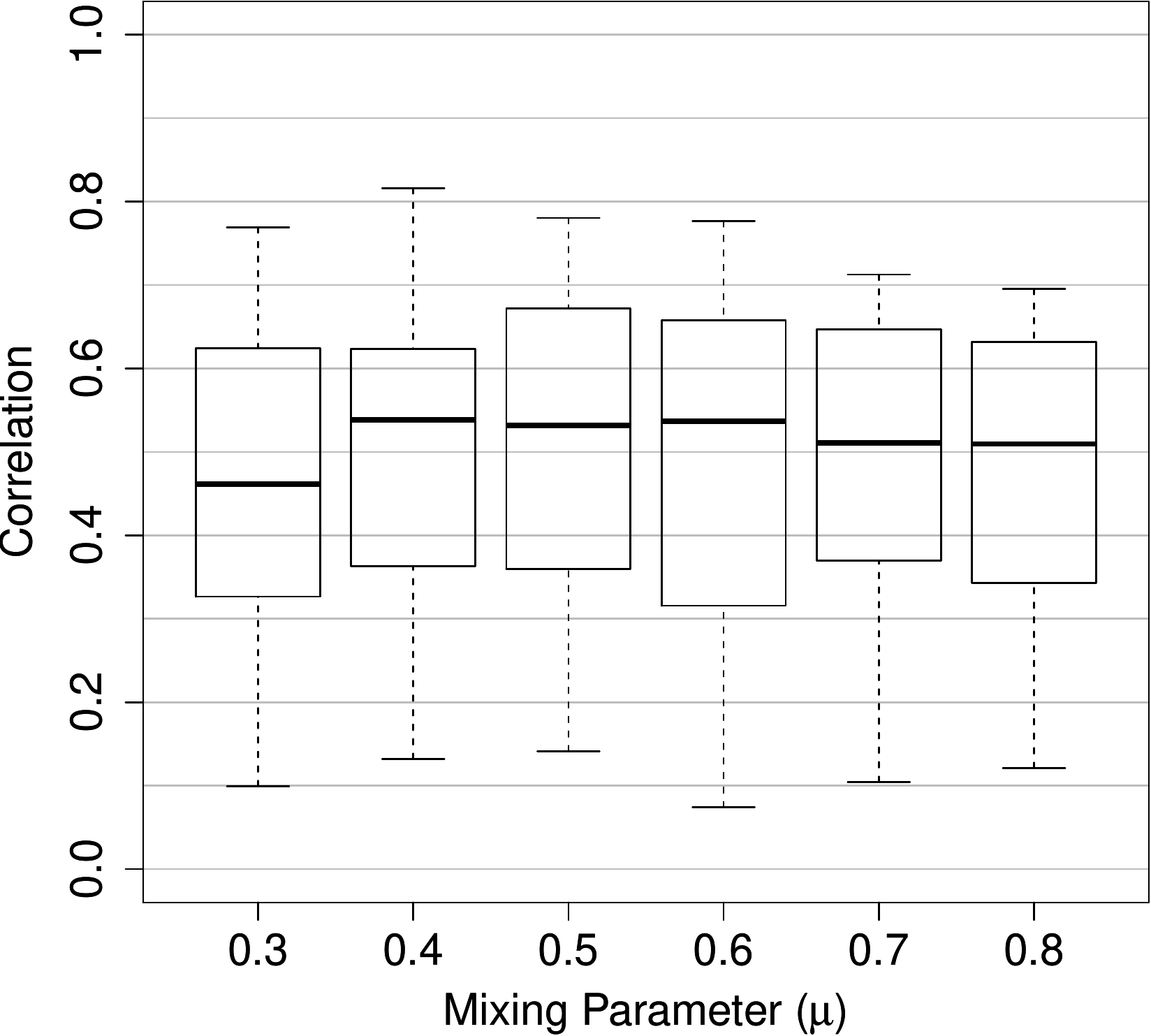}
	\end{subfigure}
	\begin{subfigure}[t]{0.24\textwidth}
		
		\includegraphics[width =\linewidth]{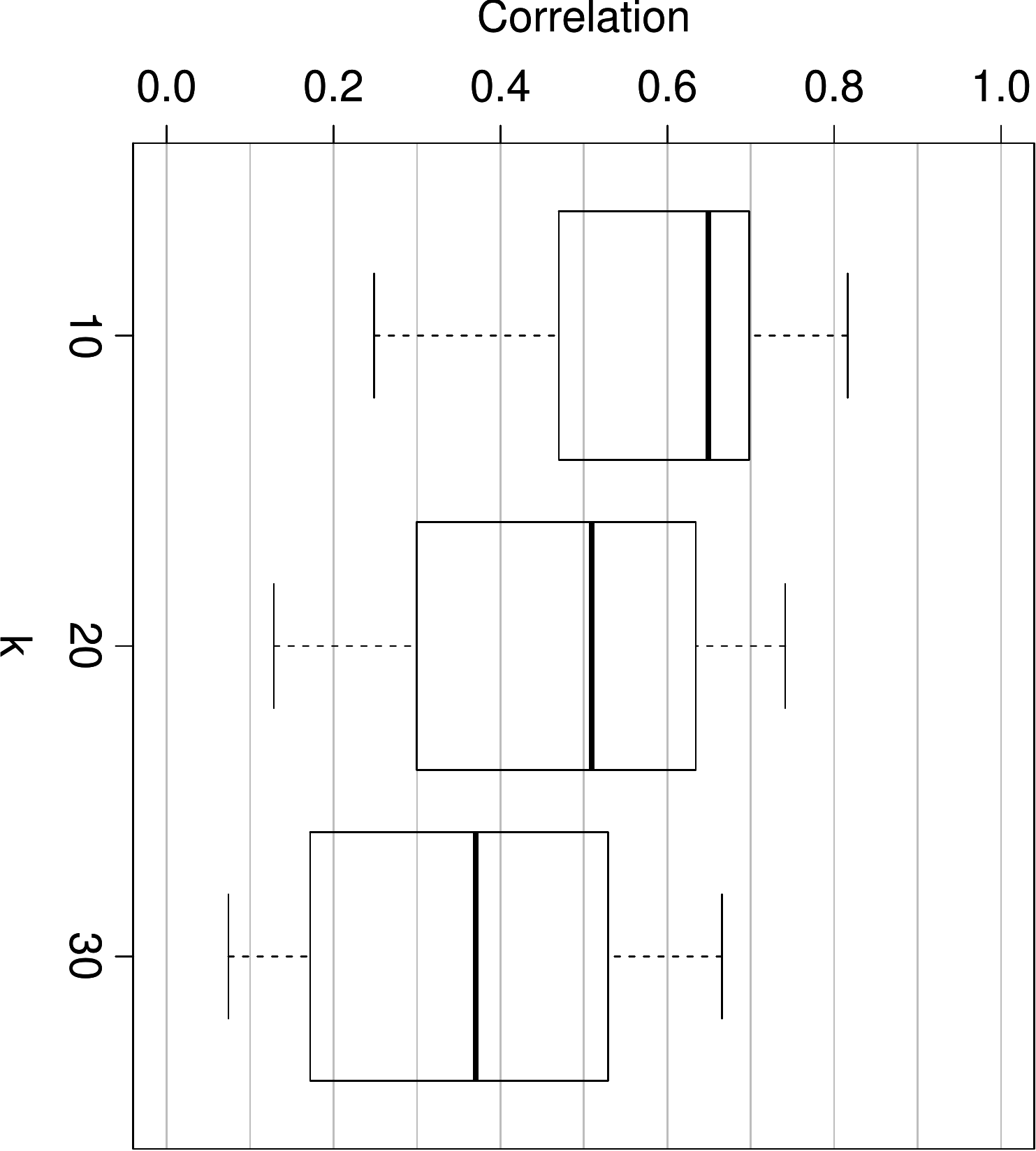}
	\end{subfigure}
	\begin{subfigure}[t]{0.24\textwidth}
		
		\includegraphics[width =\linewidth]{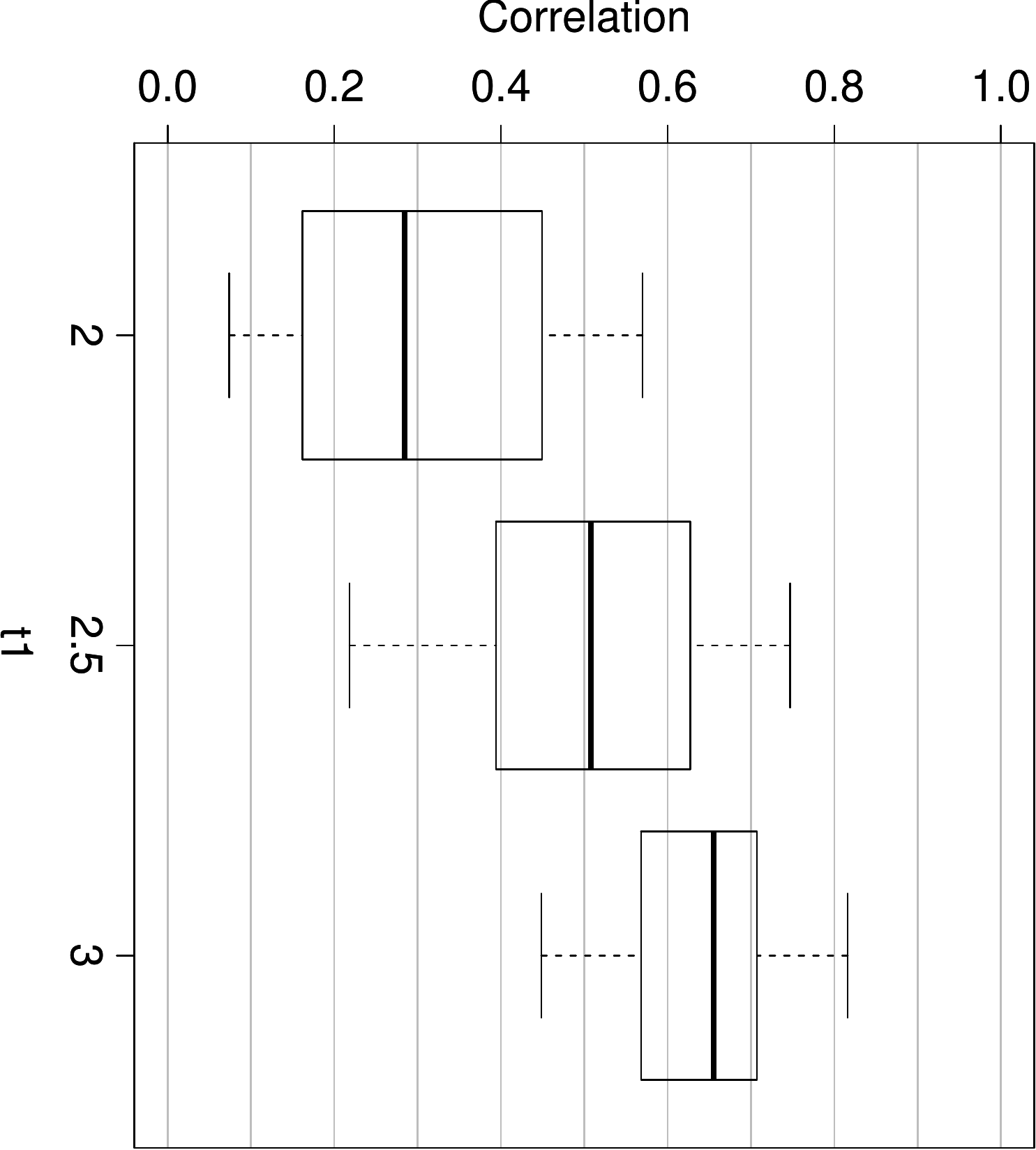}
	\end{subfigure}
		\begin{subfigure}[t]{0.24\textwidth}
		
		\includegraphics[width =\linewidth]{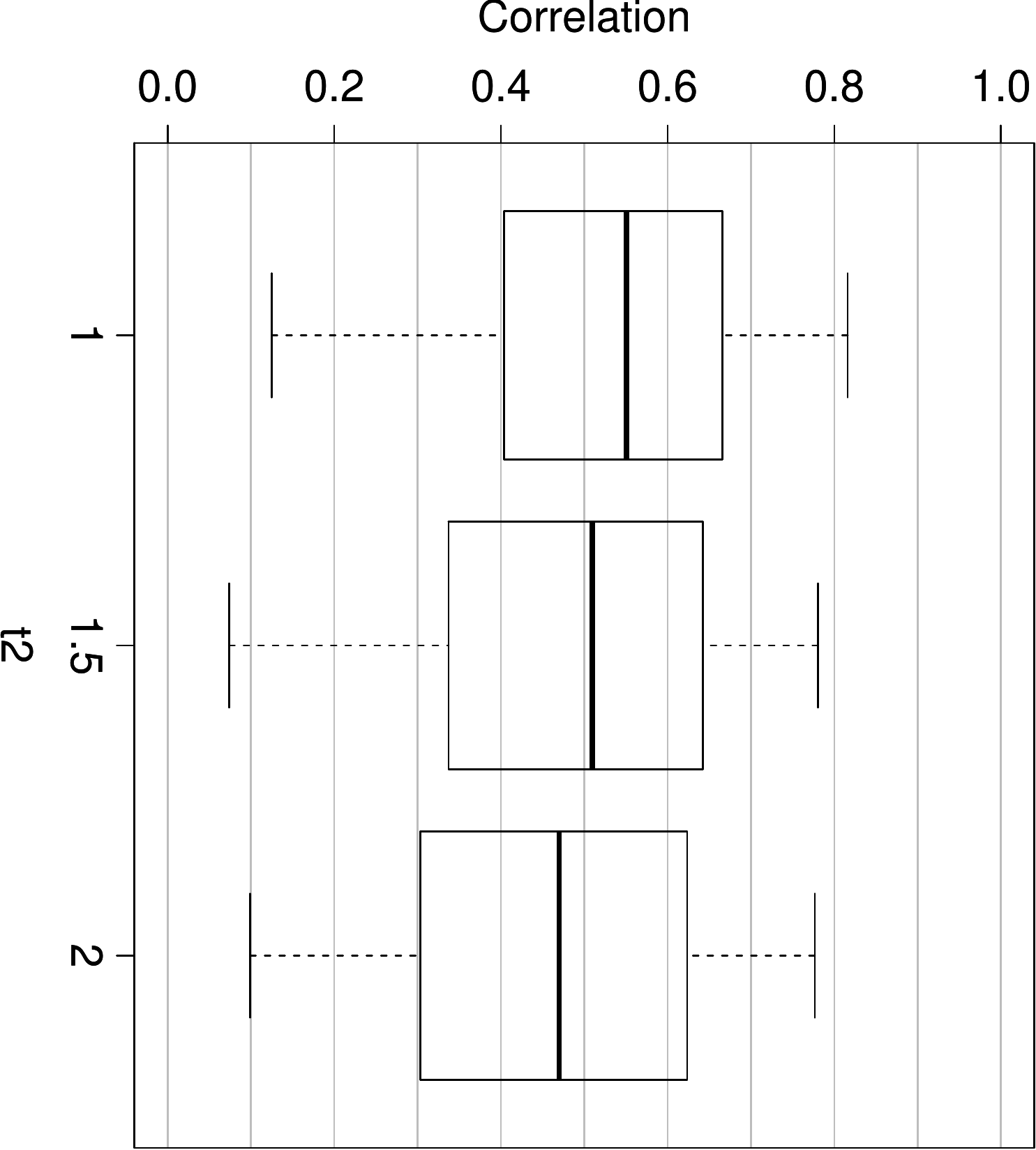}
	\end{subfigure}
\end{subfigure}
\caption{Sensitivity analysis of correlation between edge betweenness and link cohesion using LFR networks with 1000 nodes and all combinations of $\mu$ $\in$ \{0.3, 0.4, 0.5, 0.6, 0.7, 0.8\}, $k$ $\in$ \{10, 20, 30\}, $t_1$ $\in$ \{2, 2.5, 3\}, and $t_2$ $\in$ \{1, 1.5, 2\}.
}
\label{fig:corr_results}
\end{figure*}

Edge betweenness is computationally expensive to compute.  We consider the correlation coefficient between link cohesion and edge betweenness for a sweep of parameters on LFR graphs.  Our sensitivity analysis of correlation is shown in Figure~\ref{fig:corr_results}.  We observe that link cohesion and edge betweenness are positively correlated, with no significant trends or differences in correlation among the full factorial of parameter sets considered.  Hence, in cases where only a relative ranking of edges is needed, link cohesion may be useful as a surrogate score for edge betweenness.

\vspace{-1em}




\section{Conclusion}\label{sec:conclusion}

We have developed a new link cohesion metric as well as a new graph density-based pruning criteria that can be used to simplify highly connected graphs.  This method can be leveraged to improve accuracy and speed of community finding as well as other algorithms.  Similarly, the density calculation is generalizable and density-based pruning can be performed using metrics other than link cohesion.  The approach has been demonstrated using real as well as synthetic data sets with promising results.  Potential follow-on research would be to further study the relative contributions of 1-, 2-, and 3-hop link strengths to link cohesion as well as to explore other applications for density and edge pruning.

%
%

\section*{Acknowledgments}
This work was supported by internal research and development funding provided by Johns Hopkins University Applied Physics Laboratory.  Algorithms were implemented with SOCRATES \cite{savkli2014socrates} and visualized with an internal tool, Pointillist.  

\bibliographystyle{abbrv}
\bibliography{biblio.bib}

\end{document}